\documentclass[twocolumn,english,prl,floatfix,showpacs]{revtex4}
\usepackage{graphicx}
\usepackage{amssymb}

\makeatletter

\newcommand{\boldsymbol}[1]{\mbox{\boldmath $#1$}}

\usepackage{amssymb,amsmath}
\usepackage{graphicx}
\newcommand{\mnorm}[1]{
 \left\vert\kern-0.9pt\left\vert\kern-0.9pt\left\vert #1
 \right\vert\kern-0.9pt\right\vert\kern-0.9pt\right\vert}

\usepackage{babel}    
\makeatother
\begin{document}

\title{Improved transfer of quantum information using a local memory}

\author{Vittorio Giovannetti$^{1}$ and Daniel Burgarth$^{2}$}

\affiliation{$^{1}$NEST-INFM \& Scuola Normale Superiore, piazza dei Cavalieri
7, I-56126 Pisa, Italy\\
$^{2}$Department of Physics \& Astronomy, University College London,
Gower St., London WC1E 6BT, UK}

\begin{abstract}
We demonstrate that the quantum communication between two parties
can be significantly improved if the receiver is allowed to store
the received signals in a quantum memory before decoding them. In
the limit of an infinite memory, the transfer is perfect. We prove
that this scheme allows the transfer of arbitrary multipartite states
along Heisenberg chains of spin-$1/2$ particles with random coupling strengths.
\end{abstract}


\maketitle

Suppose you want to send an unknown quantum state to your friend.
Which technique should you use? Obviously you cannot just perform
a measurement and call him/her, because such a measurement would in
general only reveal very limited information about the state. Another
possibility would be to send the full physical system of the state,
but that is difficult if your state is not implemented in a mobile
medium (photons, electrons, \ldots{}) and cannot be converted to
such media easily. This is the typical situation one has to face in
solid state systems, where quantum information is usually contained
in the states of fixed objects such as quantum dots or Josephson junctions.
In this case a \emph{quantum wire} that transports states just like
optical fibers transport photons may be used. If local control (gates,
measurements) is available all along such a wire, then this state transfer is possible
via a series of swap gates or by entanglement swapping followed by
teleportation. However this scenario may be very difficult to realize
in practice. Motivated by such experimental restrictions, {\em permanently
coupled} systems without local access were suggested~\cite{BOSE},
but because of dispersion the fidelity of the transfer is in general
low. One way of improving this is by engineering specific Hamiltonians
\cite{ENG} or by coupling the system only weakly to the communicating
parties~\cite{WEAK}. 
Another approach proposed is to make use of 
gates at the sender (Alice) and the receiver (Bob) locations
and to \emph{encode} the states to be sent to yield perfect state
transfer~\cite{GAUSS,DUAL,DUAL2}.
This way the demands on the
engineering of the Hamiltonian could be relaxed. In some sense the
effort of control and engineering has been shifted to the encoding
and decoding by Alice and Bob.
Here we would like to go one step further by proposing to
make use of even more resources of Bob, i.e. to use his \emph{quantum
memory.} We will show that perfect state transfer can be achieved
using a single permanently coupled quantum chain if Bob 
possesses an
infinite quantum memory. This is achieved by \emph{swapping} the part
of the chain that Bob controls to his memory at equal time intervals.
Eventually, the whole quantum information is contained in his memory
and can be decoded by unitary operations. 
Since this happens independently of the initial state of the chain, it is an example of \emph{homogenization}~\cite{HOM} and \emph{asymptotic completeness}~\cite{AC}. The crucial difference is that in our system the memory is only interacting with Bob, and the completeness is mediated to the rest of the system through the permanent couplings. We note that with the ideas in \cite{AC} it is also possible for Bob to send messages to Alice, using the time-reversed protocol.
The
main advantage of using a memory is that - opposed to the schemes
in~\cite{BOSE,ENG,WEAK,GAUSS,DUAL,DUAL2} - Alice can send arbitrary
multi-qubit states, including complex entangled states, with a single
usage of the channel. She needs no encoding, all the work is done
by Bob. If Bob's memory is only finite, he can still use it to improve
the fidelity of the transfer substantially (the fidelity grows exponentially with the size of the memory). 
The protocol proposed 
here can 
be used to improve the performances 
of the schemes~\cite{BOSE,ENG,WEAK,GAUSS,DUAL,DUAL2}, and it works for a large
class of Hamiltonians, including Heisenberg and XY models with arbitrary
(also randomly distributed) coupling strengths. Furthermore, the 
timing 
of our protocol scales in a reasonable manner with the length of the
chain.
\begin{figure}[t]
\begin{center}\includegraphics[
  width=1.0\columnwidth]{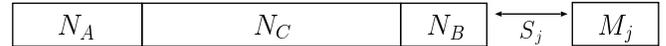}\end{center}
\caption{\label{fig:swapping}Alice and Bob control the spins $N_{A}$ and
$N_{B}$ interconnected by the spins $N_{C}.$ At time $j\tau$ Bob
performs a swap $S_{j}$ between his spins and the memory $M_{j}$.}
\end{figure}
\paragraph*{Protocol:---}
Consider a chain of spin-$1/2$ particles described by a Hamiltonian $H$ which commutes
with the total spin component $S_{z}$. The chain is assumed to be
divided in three portions $A$ (Alice), $B$ (Bob) and $C$
(the remainder of the chain, connecting Alice and Bob) containing
respectively the first $N_{A}$ spins of the chain, the last $N_{B}$
spins and the intermediate $N_{C}$ spins, and the total length of
the chain is $N=N_{A}+N_{C}+N_{B}$ (see Fig \ref{fig:swapping}).
Bob has access also  to a collection of quantum memories $M_{1},\cdots,M_{j}\cdots$
isomorphic with $B$, i.e. each having dimension equal to the dimension
$2^{N_{B}}$ of $B$. Without loosing generality it will be useful to
represent each of these memories as a non-interacting collection of
$N_{B}$ spins. The protocol goes then as follows. Suppose that at
time $t=0$ Alice prepares her spins in the (possibly unknown) input
state $|\psi\rangle_{A}$. The total state of the chain + memories
is then 
$|\psi000\rangle_{ACBM} \equiv
|\psi\rangle_{A}\otimes|0\rangle_{C}\otimes|0\rangle_{B}
\otimes|0\rangle_{M}$
where we assumed $C$, $B$ and the memories to be originally in the all-spin
down state (here $|0\rangle_{M}$ is a compact notation for the product
state $|0\rangle_{M_{1}}\otimes   
\cdots\otimes|0\rangle_{M_{j}}\cdots$).
To recover Alice message, Bob performs unitary swap operations between the
$B$ spins and the memories $M_{j}$. In particular at time $\tau>0$
Bob performs a swap between the memory $M_{1}$ 
and $B$; at
time $2\tau$ he performs a swap between $M_{2}$ and $B$;
at time $3\tau$ he performs a swap between $M_{3}$ and $B$ and so
on.
Under these hypothesis the state of the whole system after $j$ steps
is described by the unitary transformation 
\begin{eqnarray}
|\psi 000\rangle_{ACBM}\longrightarrow W_{j}|\psi 000\rangle_{ACBM}\;,
\label{mappaunitaria}
\end{eqnarray}
 where $W_{j}$ is the product of free evolutions of the chain $U\equiv\exp[-\frac{i}{\hbar}H\tau]$
and swap $S_{j}$ between the memory $M_{j}$ and $B$, i.e. 
\begin{eqnarray}
W_{j}\equiv S_{j}US_{j-1}U\cdots S_{2}US_{1}U\;.
\label{WWW}
\end{eqnarray}
For simplicity we assumed equal time intervals $\tau,$
but the generalization  to arbitrary time intervals
$\left\{ \tau_{i}\right\} _{i}$ is straightforward. The mapping $W_{j}$
preserves the total number of excitations in $A$+$B$+$C$+$M$ but tends
to decrease the number of excitations in $A$+$C$+$B$. In fact, on
one hand, the operators $U$ shuffle the spin up components of
the state around the chain $A$+$C$+$B$ while, on the other hand, the $S_{j}$
exchange the state of $B$ with the no-excitation state of the memory~$M_{j}$.
In the limit of large $j$ one expects that eventually this mechanism will provide the
transfer of $|\psi\rangle_A$ into Bob memories. 
To see how this might happen let us consider first the case $N_A,N_B=1$
where $|\psi\rangle_A$ is a generic superposition of $|0\rangle_A$ and
the spin up state $|1\rangle_A$ of $A$. 
In this context one easily verifies that
if protocol~(\ref{mappaunitaria}) stops just after the first swap, the state $|\psi\rangle_A$
can be recovered from $M_1$ 
with fidelity 
$\eta_1=| {_{ACB}\langle} 001| U | 1 00\rangle_{ACB}|^2$
identical to the transfer fidelity of Ref.~\cite{BOSE}.
If instead
protocol~(\ref{mappaunitaria}) runs up to second swap,
 $|\psi\rangle_A$ can be recovered from the state of the memories $M_1$+$M_2$ 
with fidelity 
$\eta_2=\eta_1 + | \sum_{\ell=1}^{N-1}  {_{ACB}\langle} 001  | U | \boldsymbol{\ell} \rangle_{A
CB}\langle \boldsymbol{\ell} | U | 1 00\rangle_{ACB}|^2$
which typically is already higher than the fidelity $\eta_1$ (in this expression 
$|\boldsymbol{\ell}\rangle_{ACB}$ stands for the state of  the chain with 
a single spin up component in the $\ell$-th location).
Analogously one finds that
at the $j$-th step $|\psi\rangle_A$ can be recovered from $M_1$+$\cdots$+ $M_j$ 
with a fidelity $\eta_j$ which is greater than or equal to the fidelity $\eta_{j-1}$ of the 
$(j-1)$-th step.
We claim that this a general trend which does not depend on the 
size of $N_A$ and $N_B$. In particular,
 under quite 
general hypothesis on $H$, 
we will show that  in the limit of $j\rightarrow\infty$  the input state
$|\psi\rangle_{A}$ will  be
transferred to the memories $M$ leaving the chain $A$+$B$+$C$ in the no-excitation
state $|000\rangle_{ACB}$, i.e.
 \begin{eqnarray}
\lim_{j\rightarrow\infty}W_{j}|\psi 000\rangle_{ACBM}=|000\rangle_{ACB}\otimes|\Phi(\psi)\rangle_{M}\;,\label{unounouno}\end{eqnarray}
with $|\Phi(\psi)\rangle_{M}$
being a state of $M$ which explicitly depend on the input state $|\psi\rangle_{A}$
and on~$\tau$. If the input state $|\psi\rangle_A$ does not
contain excitations 
Eq.~(\ref{unounouno})
trivially follows from the fact that for all $j$ the operator $W_{j}$
maps $|0000\rangle_{ACBM}$ into itself. For $|\psi\rangle_{A}\neq |0\rangle_{A}$
instead Eq.~(\ref{unounouno}) requires all the excitations originally
present in $A$+$C$+$B$ to move in the memory $M$ as $j$ increases. 
In our protocol, the state of $B$ is set to $|0\rangle_{B}$
at each step, so for proving Eq.~(\ref{unounouno}) it is sufficient
to show that all the excitations leave the subsystem $A$+$C$. In other
words, given the reduced density matrix 
\begin{equation}
\sigma_{AC}(j)=\textrm{Tr}_{BM}\left[W_{j} 
(|\psi 000\rangle_{ACBM}\langle\psi 000|)W_{j}^{\dag}\right]\label{eq:reduced}\end{equation}
 of $A$+$C$ at the $j$-th step of the protocol, Eq.~(\ref{unounouno})
is equivalent to requiring the following identity,\begin{eqnarray}
\lim_{j\rightarrow\infty}{_{AC}\!\langle 00}|\sigma_{AC}(j)|00\rangle_{AC}=1\;.\label{cinq1}\end{eqnarray}
Before proving this result we notice that it implies that Bob can
reliably recover Alice's messages by applying a unitary transformation on the memory only (or, alternatively, Bob could feed the memory state into another chain using the time-reversed protocol). 
In fact, since the $W_j$ are unitary operators, they describe in the limit $j\rightarrow \infty$ a unitary map from $A$ into a subspace $M_A$ of the memory of dimension $2^{N_A}$. 
The explicit form of this map depends upon the unitaries 
$U$ of Eq.~(\ref{WWW}) and can be
determined by the 
communicating parties either by knowing the chain Hamiltonian $H$ or by performing
a set of measurements prior to the transmission.

\paragraph*{Convergence:---}
We prove Eq.~(\ref{cinq1}) by showing that the probability of having
one or more excitations in $A$+$C$ at the $j$-th step of the protocol
converges to zero as $j\rightarrow\infty.$ At the beginning of the
protocol there are at most $N_{A}$ excitations in the system.
For $1\leqslant n\leqslant N_{A}$ we are interested in the probability $P_{n}(j)$
of having $n$ or more excitations in $A$+$C$ at the $j$-th step of the
protocol. This~is \begin{eqnarray}
P_{n}(j)\equiv\sum_{n^{\prime}=n}^{N_{A}}\mbox{Tr}_{AC}[\Pi_{AC}(n^{\prime})\;\sigma_{AC}(j)]\;,\label{defprobn}\end{eqnarray}
 where $\sigma_{AC}(j)$ is given by Eq.~(\ref{eq:reduced}) and $\Pi_{AC}(n^{\prime})$
are the projectors on the $\binom{N_{A}+N_{C}}{n^{\prime}}$ dimensional
Hilbert subspace of $A$+$C$ formed by the vectors with $n^{\prime}$ spins
up. An inequality for the $P_{n}(j)$ is obtained by noticing
that the total number of excitations in $A$+$C$ never increases with
$j$: this allows to upper bound $P_{n}(j+j_{1})$ with the probability
$P_{n+1}(j_{1})$ of having more than $n+1$ spins up in $A$+$C$ at the
$j_{1}$-th step plus the maximum joint probability $Q_{n}(j+j_{1},j_{1})$
of having exactly $n$ spins up at the step $j_{1}$ and maintaining
them in the next $j$ steps of the protocol, i.e.
\begin{eqnarray}
P_{n}(j_{1}+j)\leqslant P_{n+1}(j_{1})+Q_{n}(j_{1}+j,j_{1})\;.\label{define}\end{eqnarray}
The formal derivation of this rather intuitive expression 
is straightforward but tedious: we report a sketch of it in~\cite{NOTABENE}.
An expression for $Q_{n}(j_{1}+j,j_{1})$ follows by noticing
that any state of $A+C+B$ will maintain a constant number of excitations
in the chain during the  whole protocol if and only if it has no excitations
in $B$ when Bob applies the swaps $S_{j}$. For instance consider the
state $\sigma_{AC}(j_{1})$ of $A$+$C$ immediately after the $j_{1}$-th
step. According to the protocol the section $B$ is in $|0\rangle_{B}$
and the free evolution of the chain in the forthcoming time interval
is described by $U(\sigma_{AC}(j_{1})\otimes|0\rangle_{B}\langle0|)U^{\dag}$.
The probability of not loosing any excitations at step $j_{1}+1$
is then proportional to the probability that this state does not contain
excitations in $B$, i.e. \begin{eqnarray}
p_{1} & = & {_{B}\langle}0|\;\mbox{Tr}_{AC}\Big[U\;\Big(\sigma_{AC}(j_{1})\otimes|0\rangle_{B}\langle0|\Big)\; U^{\dag}\Big]\;|0\rangle_{B}\nonumber \\
 & = & \mbox{Tr}_{ACB}\Big[T\Big(\sigma_{AC}(j_{1})\otimes|0\rangle_{B}\langle0|\Big)T^{\dag}\Big]\;,\label{defp1}\end{eqnarray}
with
$T=|0\rangle_{B}\langle0|\; U$.
Moreover, if not excitations leaves  the chain at the $j_1+1$-th step, the state of $A+C+B$ 
is projected into
\begin{eqnarray}
\tilde{\sigma}_{AC}(j_{1}+1)\otimes|0\rangle_{B}\langle0|
 & = & 
\frac{1}{p_{1}}\; T\;(\sigma_{AC}(j_{1})\otimes|0\rangle_{B}\langle0|)\; T^{\dag}\;.
\nonumber 
\end{eqnarray}
 By iteration the probability that $\tilde{\sigma}_{AC}(j_{1}+1)$
will not loose excitations in the next step of the protocol is 
\begin{eqnarray}
\tilde{p}_{2} 
 & = & \mbox{Tr}_{ACB}\Big[T\Big(\tilde{\sigma}_{AC}(j_{1}+1)\otimes|0\rangle_{B}\langle0|\Big)T^{\dag}\Big]\;,\label{defp2tilde}\end{eqnarray}
 while the joint probability of not loosing excitations in the $(j_{1}+1)$-th
and in the $(j_{1}+2)$-th steps is given by $p_{2}=p_{1}\tilde{p}_{2}$,
i.e. \begin{eqnarray}
p_{2} & = & \mbox{Tr}_{ACB}\Big[T^{2}\Big(\sigma_{AC}(j_{1})\otimes|0\rangle_{B}\langle0|\Big)\left(T^{\dag}\right)^{2}\Big]\;.\label{defp2}\end{eqnarray}
 Analogously the joint probability of not loosing excitations in all
steps from $j_{1}+1$ up to $j_{1}+j$ is  equal to \begin{eqnarray}
p_{j} & = & \mbox{Tr}_{ACB}\Big[T^{j}\Big(\sigma_{AC}(j_{1})\otimes|0\rangle_{B}\langle0|\Big)\left(T^{\dag}\right)^{j}\Big]\;.\end{eqnarray}
 The quantity $Q_{n}(j_{1}+j,j_{1})$ can now be computed by assuming
$\sigma_{AC}(j_{1})$ to have exactly $n$ excitations and maximizing
$p_{j}$ with respect to such a choice, i.e. \begin{eqnarray}
&&Q_{n}(j_{1}+j,j_{1})\nonumber 
  =  \max_{|\phi_{n}\rangle_{AC}}\;\mbox{Tr}_{ACB}
\Big[T^{j}\Big(|\phi_{n}\rangle_{AC}\langle\phi_{n}| \nonumber \\
&&\!\otimes|0\rangle_{B}\langle0|\Big)\left(T^{\dag}\right)^{j}\Big]
 =  \max_{|\phi_{n}\rangle_{AC}}\;\Vert T^{j}\;\big(|\phi_{n}0\rangle_{ACB}\big)\Vert^{2},
\label{FRA}\end{eqnarray}
 where $|\phi_{n}\rangle_{AC}$ is a generic state of $A$+$C$ with $n$ excitations
and \mbox{$|\phi_{n}\;0\rangle_{ACB}\equiv|\phi_{n}\rangle_{AC}\otimes|0\rangle_{B}$}.
Notice that by exploiting the convexity of mixed states, the maximization
in Eq.~(\ref{FRA}) 
has been performed only on pure states. For $n=0$ it is trivial to see that $Q_{0}(j_{1}+j,j_{1})=1$
for all $j$ and $j_1$.
We will show now that for
$n\geqslant 1$ and $j_1\geqslant 0$, one has instead
\begin{eqnarray}
\lim_{j\rightarrow\infty}Q_{n}(j_{1}+j,j_{1})=0\;.\label{deftesi}\end{eqnarray}
 Because the operator $T$ conserves the number of excitations, 
we get \begin{equation}
\Vert T^{j}|\phi_{n}\;0\rangle_{ACB}\Vert^{2}=\Vert T_{n}^{j}|\phi_{n}\;0\rangle_{ACB}\Vert^{2},\label{eq:restriction}\end{equation}
where $T_{n}=\Pi_{ACB}(n)\: T\:\Pi_{ACB}(n)$ is the restriction of
$T$ to the subspace with $n$ excitations. Eq.~(\ref{eq:restriction})
converges to zero for all $\phi_{n}$ iff the spectral radius $\rho(T_{n})$
of $T$ is smaller than one~\cite{HORN}. Since $T_{n}$ is the product
of a projector and a unitary operator, it is easy to see that this
is the case~\cite{DUAL2} iff there exists no common eigenstate of
$|0\rangle_{B}\langle0|$ and $U_{n}=\Pi_{ACB}(n)\: U\:\Pi_{ACB}(n)$.
Because $U_{n}=\sum\exp\left(-iE_{n}\tau\right)|E_{n}\rangle\langle E_{n}|,$
where $|E_{n}\rangle$ are the eigenstates of the Hamiltonian $H$
with exactly $n$ excitations, it is always possible
to find a choice for the interval $\tau$ such that Eq.~(\ref{deftesi})
holds, as long as given $n\geqslant 1$ there are no eigenstates $|E_{n}\rangle$ of factorizing
form with $|0\rangle_{B}$, i.e. \begin{equation}
\nexists\:|\lambda_{n}\rangle_{AC}:\quad H|\lambda_{n}\rangle_{AC}\otimes|0\rangle_{B}=E|\lambda_{n}\rangle_{AC}\otimes|0\rangle_{B}.\label{eq:condition}\end{equation}
 Under this condition Eq.~(\ref{define}) implies that for
any $\delta_{1}>0$, there exists a sufficiently big $J_{1}$ such
that for all $j>J_{1}$ one has 
$P_{n}(j_{1}+j)\leqslant P_{n+1}(j_{1})+\delta_{1}$.
Reiterating this with respect to $n$ one can show  that given $\delta>0$
there is $J$ such that for all $j>J$\begin{eqnarray}
P_{n}(j_{1}+j)\leqslant P_{N_{A}}(j_{1})+\delta\;,\label{define2}\end{eqnarray}
 where $N_{A}$ is the maximum number of spin up Alice can introduce
in $A$. From our definitions the quantity $P_{N_{A}}(j_{1})$ is the
probability of having $N_{A}$ spins up in $A$+$C$+$B$ at the $j_{1}$-th
step. This quantity cannot be greater than $Q_{N_{A}}(j_{1},0)$ of
Eq.~(\ref{FRA}). But according to Eq.~(\ref{deftesi}) this nullifies
in the limit $j_{1}\rightarrow\infty$. Therefore for $n\geqslant1$ one has
$\lim_{j\rightarrow\infty}P_{n}(j)=0$
which gives the thesis.

\paragraph*{Nearest-neighbor interactions:---}
 The requirement~(\ref{eq:condition}) is quite general, and does not require
any particular constraint on  the topology of the system (e.g. it does not need
to be a chain). However in the following we will focus on the special case of linear open
chains showing that~(\ref{eq:condition}) is always satisfied if they
 a) conserve the number of excitations and b) are connected
by nearest-neighbor 
exchange terms. This includes
the randomly coupled chains considered in~\cite{DUAL}.
Consider in fact one of such chain and 
assume by contradiction it has an eigenvector $|E_n\rangle$
which falsifies  Eq.~(\ref{eq:condition}) for some $n\geqslant 1$. 
Such an eigenstate
can be written as
\begin{equation}
|E_{n}\rangle=a|\mu_{n}\rangle_{AC}\otimes|0\rangle_{B}+b|\bar{\mu}_{n}\rangle_{AC}\otimes|0\rangle_{B},\label{eq:eig}\end{equation}
where $a$ and $b$ are complex coefficients and where the spin just
before the section $B$ (with position $N_{A}+N_{C}$) is in the state
$|0\rangle$ for $|\mu_{n}\rangle_{AC}$ and in the state $|1\rangle$
for $|\bar{\mu}_{n}\rangle_{AC}.$ Since the interaction between this
spin and the first spin of section $B$ includes an exchange term, then
the action of $H$ on the second term of~(\ref{eq:eig}) yields exactly one state which
contains an excitation in the sector $B$ which cannot be compensated by the
action of $H$ on the first term of~(\ref{eq:eig}).
But by assumption $|E_{n}\rangle$
is an eigenstate of $H$, so we conclude that $b=0.$ This argument can be
repeated for the second last spin of section $C$, the third last spin,
and so on, to finally yield $|E_{n}\rangle=|0\rangle_{ACB}$, as long
as all the nearest neighbor interactions contain exchange parts. This
leads to a contradiction for $n\geqslant 1$.
\paragraph*{Time-scale:---}
As we have shown above, the communicating parties can achieve perfect
state transfer in the limit of infinite time and an infinitely large
memory space. However in practice, Bob's resources and time will be
limited. If the protocol stops after $j$ operations, how does the
fidelity depend on the number of qubits $N_{A}$ being transferred,
and on the total length of the chain? This question is clearly strongly
depending on the specific Hamiltonian of the chain. For example, in
the case of engineered couplings~\cite{ENG}, a \emph{single} swap
operation would already suffice. We would like to keep the argument
in this section as general as possible to find a rough estimate of
the fidelity based on statistical arguments. If the system has some
special symmetries, the fidelity may be much higher, as in the case
of engineered couplings, or may also be much lower, but in practice
these cases are extremely unlikely.

Since the transfer of spin-down components occurs naturally in our model, one may argue that
the worst case scenario is when Alice
wants to send the state $|1\:1\:\ldots\:1\rangle_{A}$.
After an initial time $T_{e}$ that it takes excitations to travel across the chain, we expect
that the $N_A$ excitations originally at Alice's site are distributed
with an average number of $N_{A}/N$ excitations per site.
On average, Bob's region of the chain should therefore contain
$N_{B}N_{A}/N$
 excitations. Of course the expectation value of the number of excitations
is a strongly fluctuating function of time. However in a slightly
modified protocol with optimized swapping times $\left\{ \tau_{i}\right\} _{i},$
it should be easy to find a swapping time $\tau_{1}\in\left[0,T_{e}\right]$
such that after performing the swap operation, there are on average
$N_{1}=\left(1-N_{B}/N\right)N_{A}$
excitations left which remain in the part $A$+$C$ of the chain. After
another time of the order of $T_{e},$ they will be spread along the
whole chain again, with $N_{B}N_{1}/N$ being the average number in
Bob's section. More generally, after a time $t \approx j T_{e}$ the average 
number of excitations in the system after $j$ swap
should be of the order
$N_{j}=\left(1-N_{B}/N\right)^{j}N_{A}$
 (we have confirmed this estimate numerically for short Heisenberg spin chains).
The fidelity
$F$ of the state transfer is lower bounded by the probability 
of having no excitations in the chain $A$+$C$+$B$. For $N_{j}\leqslant 1$ we
can lower bound this quantity 
by $1-N_{j}$. Thus for
large $j$ one has
$F\geqslant 1-\left(1-N_{B}/N\right)^{j}N_{A}$.
Replacing $j\simeq t/T_e$ and taking  the limit $N>>1$ the above inequality
shows that the fidelity $F$ can be reached after a time
$t \approx N T_e (\ln N_{A}+|\ln (1-F ) | )/N_{B}$.
In
translationally invariant systems
 the group velocity is typically independent of the length $N$ 
of the chain. Therefore in these systems $T_e$ is scaling linearly with $N$~\cite{BOSE}
and the above equation shows that $t$ scales quadratically with $N$.
 A special case of this expression with $N_{A}=N_{B}=1$ and
$1-F$ corresponding to a probability of failure was already considered
in the conclusive dual rail schemes~\cite{DUAL}. 
From the above analysis it follows that the size of Bob's region can make
the transfer quicker, and that the time-scale only depends logarithmically
on the amount of qubits that Alice wants to send. It is therefore
more efficient to send many qubits at once rather than repeating the
protocol.

\paragraph*{Conclusions:---}
We have shown that the usage of the quantum memory of the receiver
can strongly increase the fidelity of quantum state transfer with
permanently coupled quantum chains. In the limit of an infinite memory,
the transfer is perfect. Furthermore this scheme allows to send arbitrary
multipartite states rather than just single qubit states.

\paragraph*{Acknowledgments:---}
DB acknowledges
the support of the UK EPSRC through the
Grant Nr. GR/S62796/01. DB would like to thank Sougato Bose for fruitful
discussions.


\begin{thebibliography}{1}
\bibitem{BOSE}S. Bose, Phys. Rev. Lett. \textbf{91}, 207901 (2003).
\bibitem{ENG}M. Christandl, {\em et al.},
Phys. Rev. Lett.
\textbf{92}, 187902 (2004); G. M. Nikolopoulos, D. Petrosyan and P.
Lambropoulos, J. Phys.: Condens. Matter \textbf{16}, 4991 (2004);
M. H. Yung and S. Bose, Phys. Rev. A \textbf{71}, 032310 (2005); P. Karbach
and J. Stolze, {\em ibid.} {\bf 72}, 030301 (2005).
\bibitem{WEAK}M. B. Plenio and F. L. Semiao, New. J. Phys. \textbf{7}, 73 (2005);
Y. Li, {\em et al.}, 
 Phys. Rev. A \textbf{71},
022301 (2005); A. Wojcik, {\em et al.}, 
{\em ibid.} {\bf 72}, 034303 (2005).
\bibitem{GAUSS}T. J. Osborne and N. Linden, Phys. Rev. A \textbf{69}, 052315 (2004);
V. Giovannetti and R. Fazio,  {\em ibid.} {\bf 71}, 032314 (2005);
H. L. Haselgrove, quant-ph/0404152. 
\bibitem{DUAL}D. Burgarth and S. Bose, Phys. Rev. A \textbf{71}, 052315 (2005);
 New. J. Phys. \textbf{7}, 135 (2005).
\bibitem{DUAL2}D. Burgarth, V. Giovannetti, and S. Bose, 
J. Phys. A: Math. Gen. {\bf 38} 6793 (2005).
\bibitem{HOM} M. Ziman {\em et al.}, Phys. Rev. A \textbf{65},
042105 (2002).
\bibitem{AC}T. Wellens {\em et al.}, Phys. Rev. Lett. \textbf{85}, 3361 (2000)
\bibitem{NOTABENE}The key ingredient to derive~(\ref{define}) from~(\ref{defprobn})
is  the relation $\sigma_{AC}(j+1)
= \mbox{Tr}_B[ U( \sigma_{AC}(j)\otimes |0\rangle_B\langle 0| )U^\dag ]$
which connects the reduced density matrices of $A+C$ at successive steps
of the protocol. Use this to express $\sigma_{AC}(j_1 +j)$ of $P_{n}(j_{1}+j)$
in terms of $\sigma_{AC}(j_1)$ and, for $D=B, AB$, introduce the decompositions 
$\openone_D= \sum_{n^\prime=0}^\infty \Pi_{D}(n^{\prime})$, with
$\Pi_{D}(n^\prime)$ defined as in Eq.~(\ref{defprobn}) and $\openone_{D}$
being the identity operator of $D$.
Since $U$ preserves the number of excitation in $A+C+B$, 
the expression~(\ref{defprobn}) of  $P_{n}(j_{1}+j)$ separates in
two contributions: the first includes only terms of $\Pi_{AC}(n^\prime) \sigma_{AC}(j_1)
\Pi_{AC}(n^\prime)$ with $n^\prime \geqslant n+1$, and can be upper bounded
by $P_{j+1}(n)$; the second includes only $\Pi_{AC}(n) \sigma_{AC}(j_1)
\Pi_{AC}(n)$ and can be upper bounded by $Q_{n}(j_{1}+j,j_{1})$ of Eq.~(\ref{FRA}).
\bibitem{HORN}R. A. Horn and C. R. Johnson, Matrix Analysis, Cambridge University
Press, 1990.
\end{thebibliography}
\end{document}